\documentclass{iau}
\usepackage{graphicx}
\usepackage{subfigure}

\title[Subtle flickering in Cepheids] %% give here short title %%
{Subtle flickering in Cepheids: \\ {\it Kepler} and {\it MOST}}

\author[N.R.~Evans et al.]   %% give here short author list %%
{Nancy Remage Evans$^1$,
Robert Szab\'o$^2$, Laszlo Szabados$^2$,\\ Aliz  Derekas$^2$, Jaymie
Matthews$^3$, Chris Cameron$^3$,\\ \and the {\it MOST} Team}

\affiliation{$^1$SAO, MS 4, 60 Garden St., Cambridge, MA 02138, USA\\
email: {\tt nevans@cfa.harvard.edu}\\[\affilskip]
$^2$MTA CSFK Konkoly Thege Miklos ut 15-17, H-1121 Budapest Hungary\\[\affilskip]
$^3$Dept. of Physics and Astronomy, Univ. of British Columbia,
Vancouver, BC, Canada}

\pubyear{2014}
\volume{301}  %% insert here IAU Symposium No.
\pagerange{1--4}
\jname{Precision Asteroseismology}
\editors{J.A. Guzik, W.J. Chaplin, G. Handler \& A. Pigulski, eds.}
\begin{document}

\maketitle

\begin{abstract}
Fundamental mode classical Cepheids have light curves which repeat
accurately  enough that we 
can watch them evolve (change period). The new level of accuracy  and 
quantity of data with the {\it Kepler} and {\it MOST} satellites
probes this further.
An  intriguing result was found in the long time-series of Kepler data
for V1154 Cyg the one classical Cepheid (fundamental mode, P = 4.9$^d$) in the
field, which has short term changes in period ($\simeq$20 minutes),
correlated for $\simeq$10 cycles (period jitter).  To follow this up, we
obtained a month long series of observations of the fundamental mode 
Cepheid RT Aur and the first overtone pulsator SZ Tau.  RT Aur shows 
the traditional strict repetition of the light curve, with the 
Fourier amplitude ratio $R_1/R_2$ remaining nearly constant. 
 The light curve of  SZ Tau, on the other hand, fluctuates 
in amplitude ratio  at the level of approximately 50\%.  Furthermore 
prewhitening the RT Aur data with 10 frequencies reduces the Fourier 
spectrum to noise.  For SZ Tau, considerable  power is left after 
this prewhitening in a complicated variety of frequencies.  
%% add here a maximum of 10 keywords, to be taken form the file <Keywords.txt>
\end{abstract}

\firstsection % if your document starts with a section,
              % remove some space above using this command.
\section{Introduction}

The quality and quantity of satellite data is revealing subtle
features in Cepheid pulsation.  Specifically, we discuss here recent
findings from two long series of observations by the {\it Kepler} and {\it
  MOST} satellites.  

\section{{\it Kepler} observations}

There is only one classical Cepheid (V1154 Cyg) in the {\it Kepler} field, and
it is a fundamental mode pulsator with a period of 4.9$^d$ .  Analysis of the {\it Kepler} data by 
\cite[Derekas et al. (2012)]{Derekas etal12} 
 found small cycle to cycle variations of the period.
Typically the period excursions might be 20 minutes and last up to 15
cycles before the sign of the period change ($\dot{P}$) reverses.

\section{{\it MOST} observations}

\begin{figure}
% \vspace*{-2.0 cm}
\begin{center}
 \includegraphics[width=2.0in]{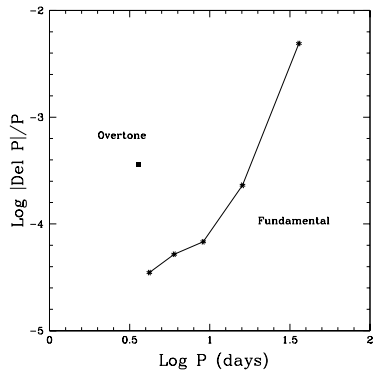} 
% \vspace*{-1.0 cm}
 \caption{Fractional period change (absolute value)  as a function of
   the log of the   pulsation period in days. Fundamental
   mode pulsators (binned) are denoted by asterisks; overtone
   pulsators  are a filled square.}
   \label{fig1}
\end{center}
\end{figure}

To look further for the small fluctuations made visible by the long
continuous strings of accurate satellite data, we observed two
Cepheids with the {\it MOST} satellite (\cite[Walker et
  al.~2003]{Walker etal03}, \cite[Matthews et al.~2004]{Matthews etal04}).
Motivation for the structure of the observation request is provided by
known period changes in Cepheids (Fig.~\ref{fig1}).  The data taken from
\cite[Szabados (1983)]{Szabados83}
 show the well known increase in period fluctuations
with period or luminosity of the Cepheid.  This is consistent with
period changes ($\dot{P}$) determined by evolution through the instability
strip, with more luminous stars evolving more quickly.  The exception
to the trend, however, is the group of  Cepheids pulsating in an
overtone mode 
which apparently show unusually large period jitter rather than the
smooth and continuous evolution of fundamental mode pulsators.  

\begin{figure}
%\begin{subfigure}[b]{0.5\textwidth}
% \vspace*{-2.0 cm}
\begin{center}
\subfigure{\includegraphics[width=2.0in]{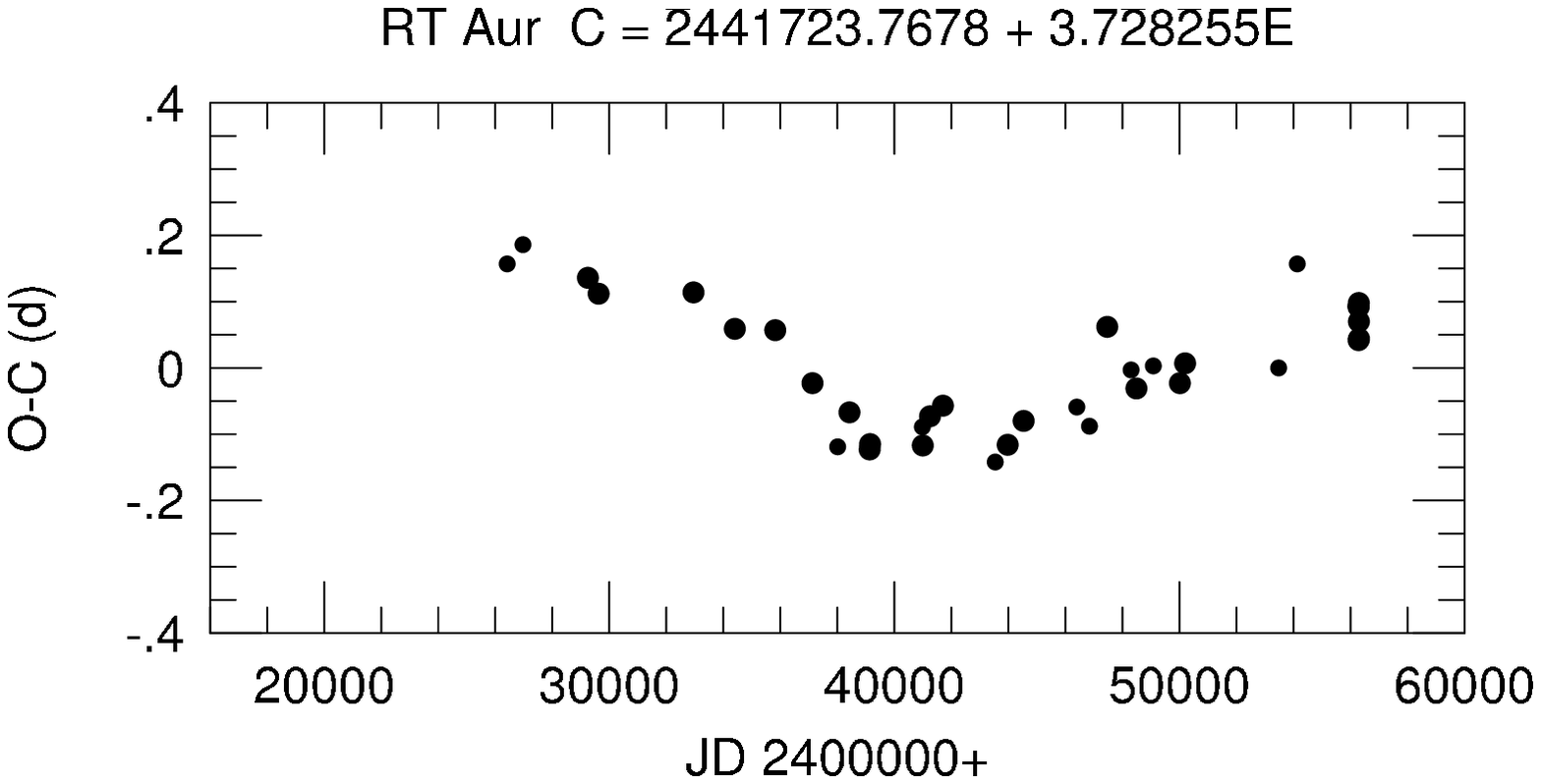}}
% \end{subfigure}%
% \vspace*{-1.0 cm}
%\begin{subfigure}[b]{0.5\textwidth}
\subfigure{\includegraphics[width=2.0in]{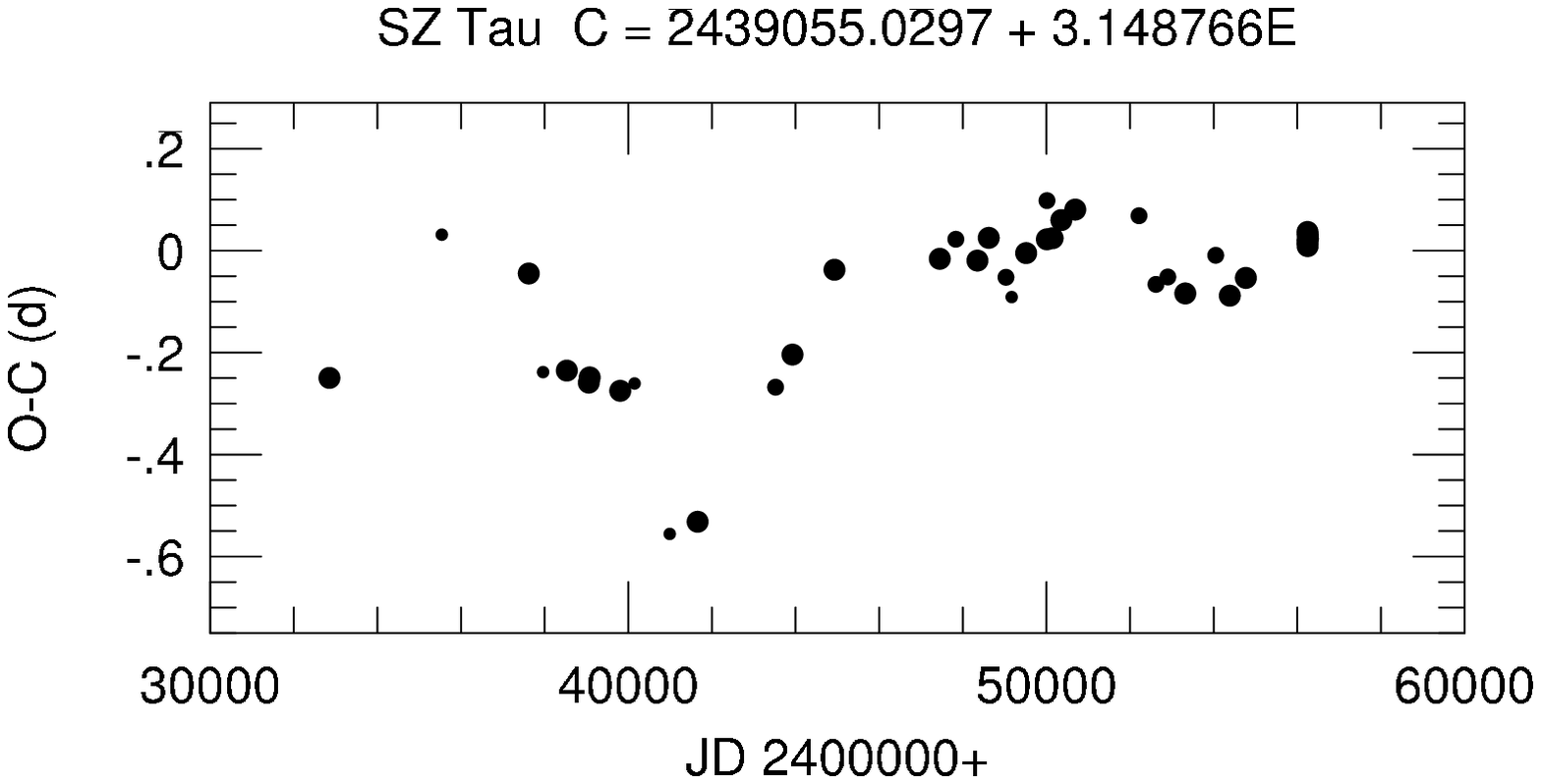}}
%\includegraphics[width=2.0in]{evans.fig2b.eps} 
%\end{subfigure}%
 \caption{The long-term period variations (O$-$C diagrams: O$-$C in
   days as a function of Julian date - 2,400,000) for RT~Aur
   (left) and SZ~Tau (right). 
As is typical of their pulsation modes,
 the variation for RT~Aur is smooth, where for SZ~Tau it is both positive and
 negative and also constant for a long period. 
Symbol size denotes the significance of the O-C value.}
   \label{fig2}
\end{center}
\end{figure}

For the {\it MOST} observations we wanted to contrast the behavior of a
fundamental mode pulsator (RT Aur) with a first overtone pulsator (SZ Tau). 
The background information on $\dot{P}$ is shown in Fig.~\ref{fig2}.  For RT Aur the
period change O$-$C (observed minus computed) diagram has a decrease
followed by an increase, which is typically fitted with 
 a parabola.  For SZ Tau, the O$-$C diagram shows both
increasing and decreasing periods, as well as about 10\,000 days when
it remains constant.
  
\begin{figure}
%\begin{subfigure}[b]{0.5\textwidth}
% \vspace*{-2.0 cm}
\begin{center}
\subfigure{\includegraphics[width=1.5in,angle=270]{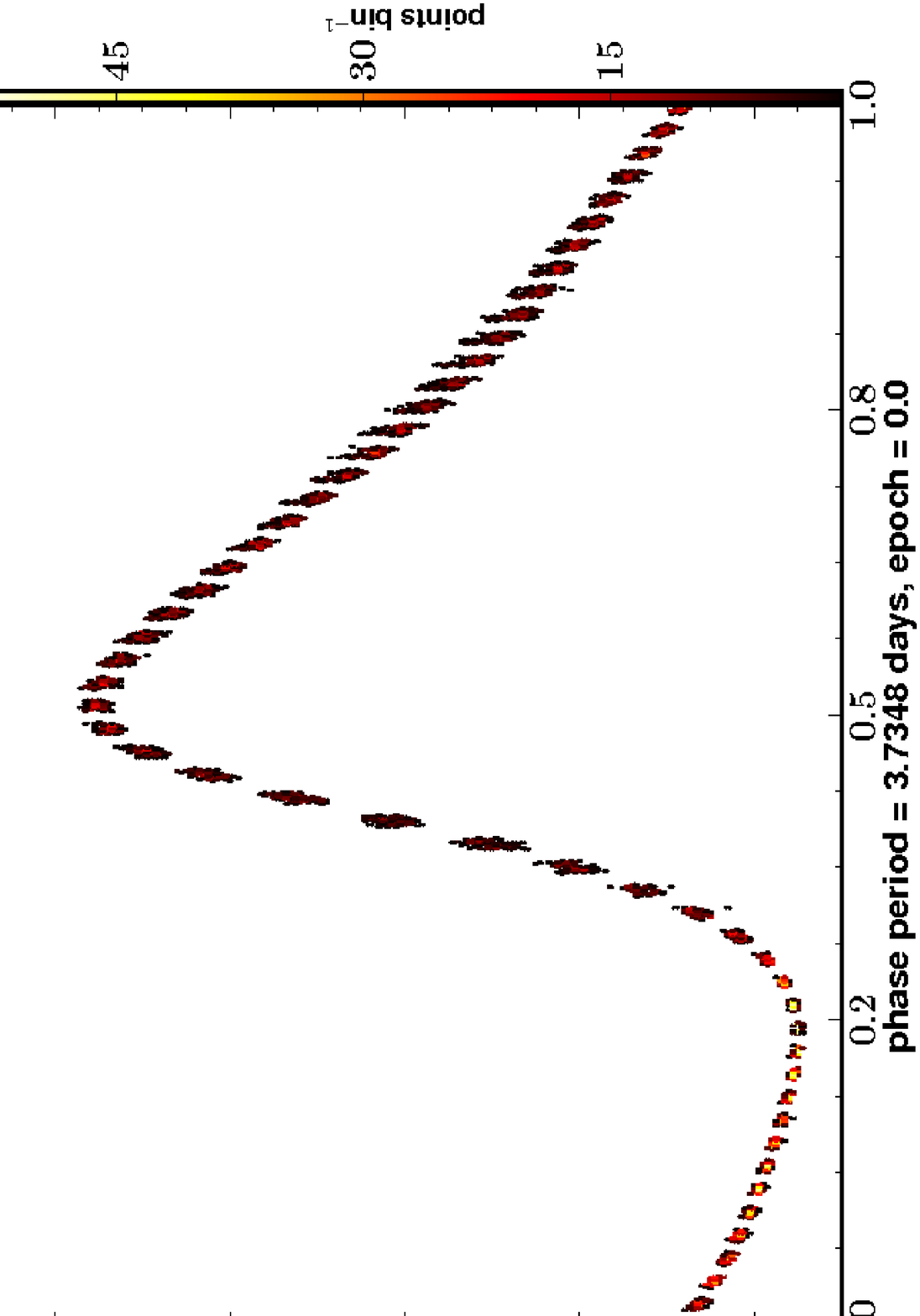}}
% \end{subfigure}%
% \vspace*{-1.0 cm}
%\begin{subfigure}[b]{0.5\textwidth}
\subfigure{\includegraphics[width=1.5in,angle=270]{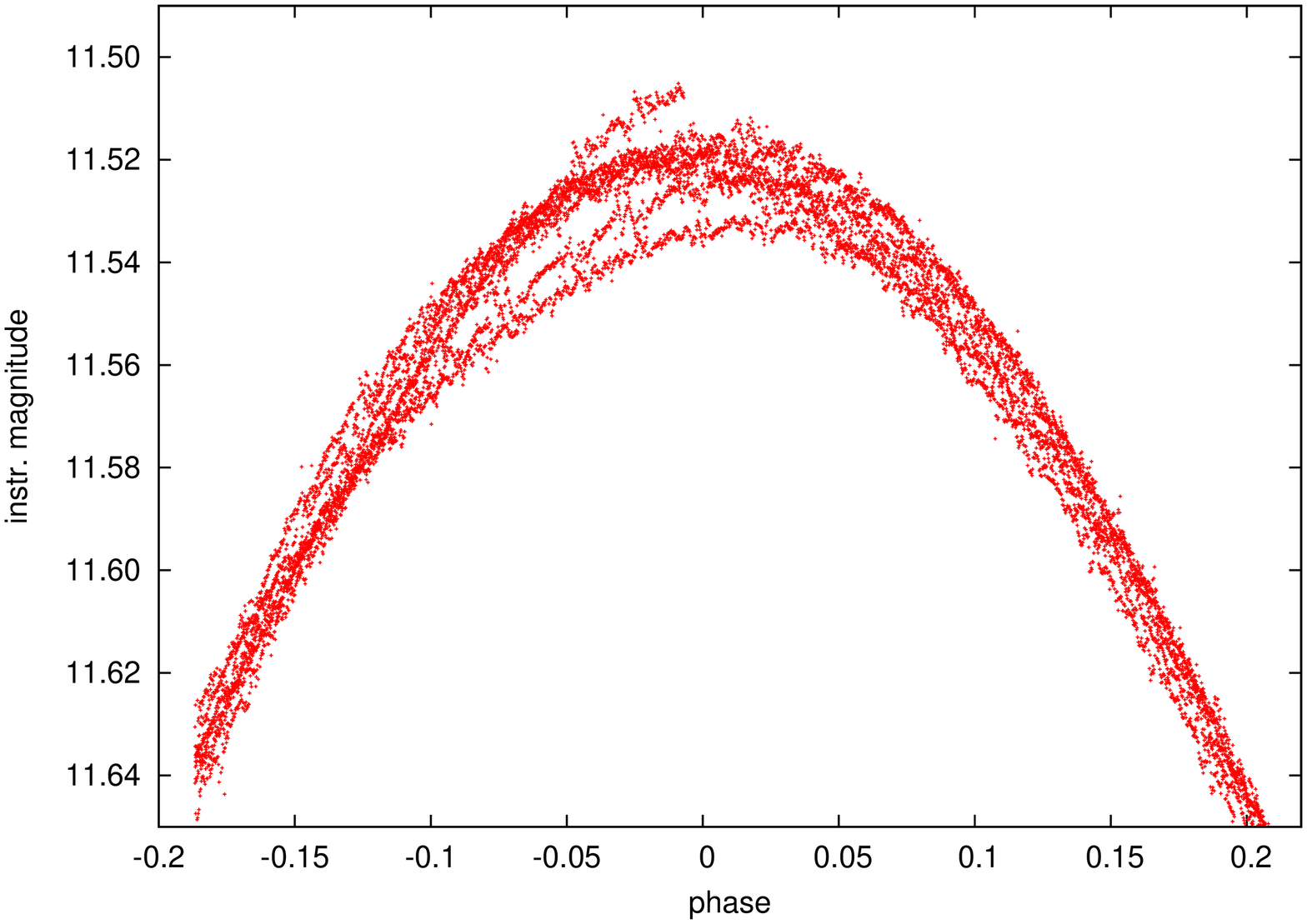}} 
%\end{subfigure}%
 \caption{The {\it MOST} light curves (magnitudes as a function of
pulsation   phase) for RT Aur (left) and SZ Tau
   (right).  The RT Aur figure also shows points per bin on the
right side scale. The RT Aur observations were
 interleaved with another target, resulting in the gaps in the data.
 Only the maximum light is shown for SZ Tau, emphasizing the variation
 in brightness between cycles. }
   \label{fig3}
\end{center}
\end{figure}

The {\it MOST} observations (Evans et al., in preparation) phased for pulsation period
are shown in Fig.~\ref{fig3}.  For RT Aur, the observations were interleaved
with another target, resulting in gaps in the light curve.  However,
the light curve repeats very precisely from cycle to cycle.  For SZ
Tau, the overtone pulsator, only the maximum of the phased light curve
is shown.  This emphasizes the fact that there are variations in
maximum brightness from cycle to cycle (also easily visible at minimum
light). A small instrumental signal (differing earthshine through
the satellite orbit) is seen in Fig.~\ref{fig3}b, which is being removed
through additional processing.  

\begin{figure}
%\begin{subfigure}[b]{0.5\textwidth}
% \vspace*{-2.0 cm}
\begin{center}
\subfigure{\includegraphics[width=1.6in,angle=270]{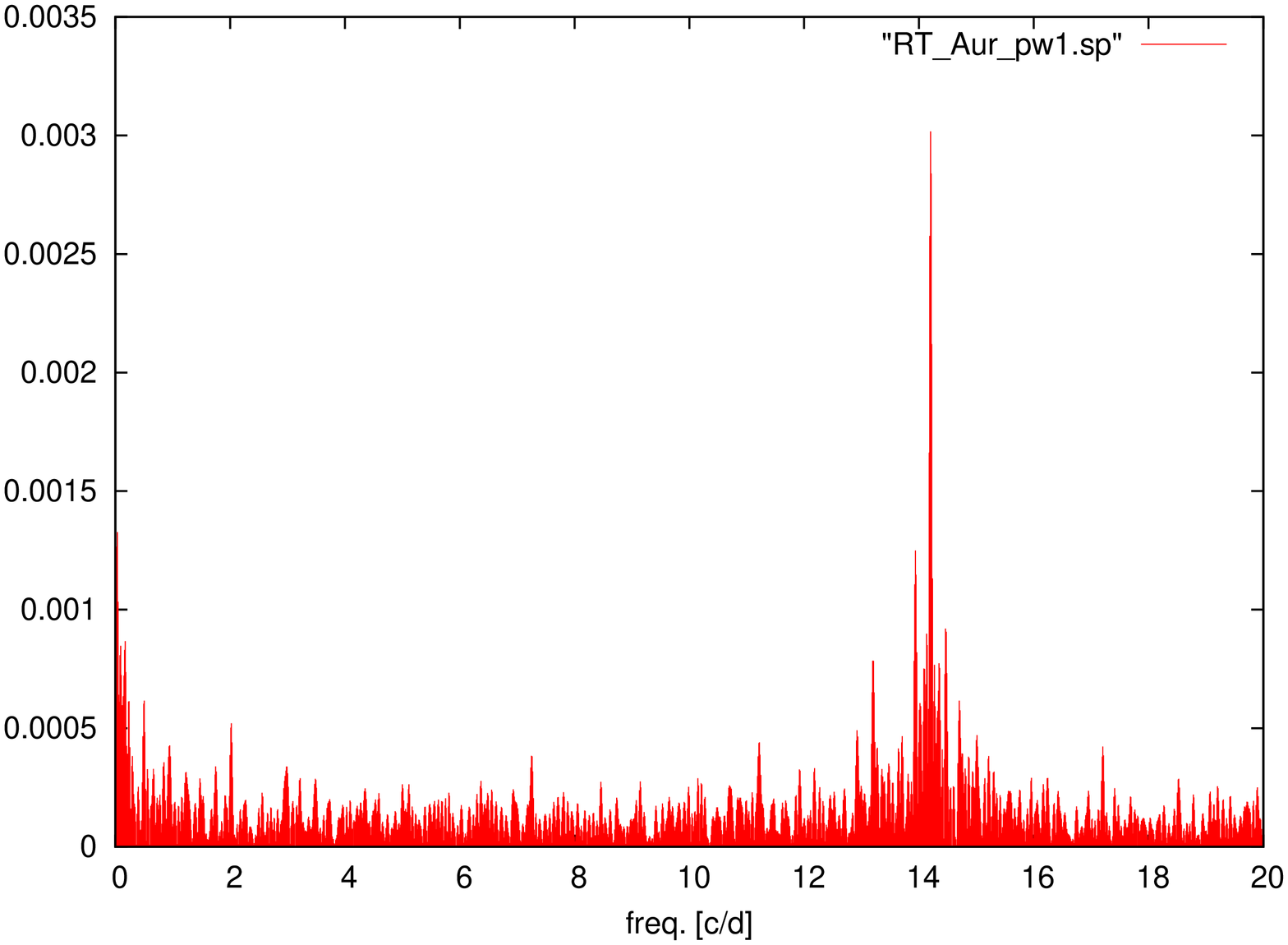}}
% \end{subfigure}%
% \vspace*{-1.0 cm}
%\begin{subfigure}[b]{0.5\textwidth}
\subfigure{\includegraphics[width=1.6in,angle=270]{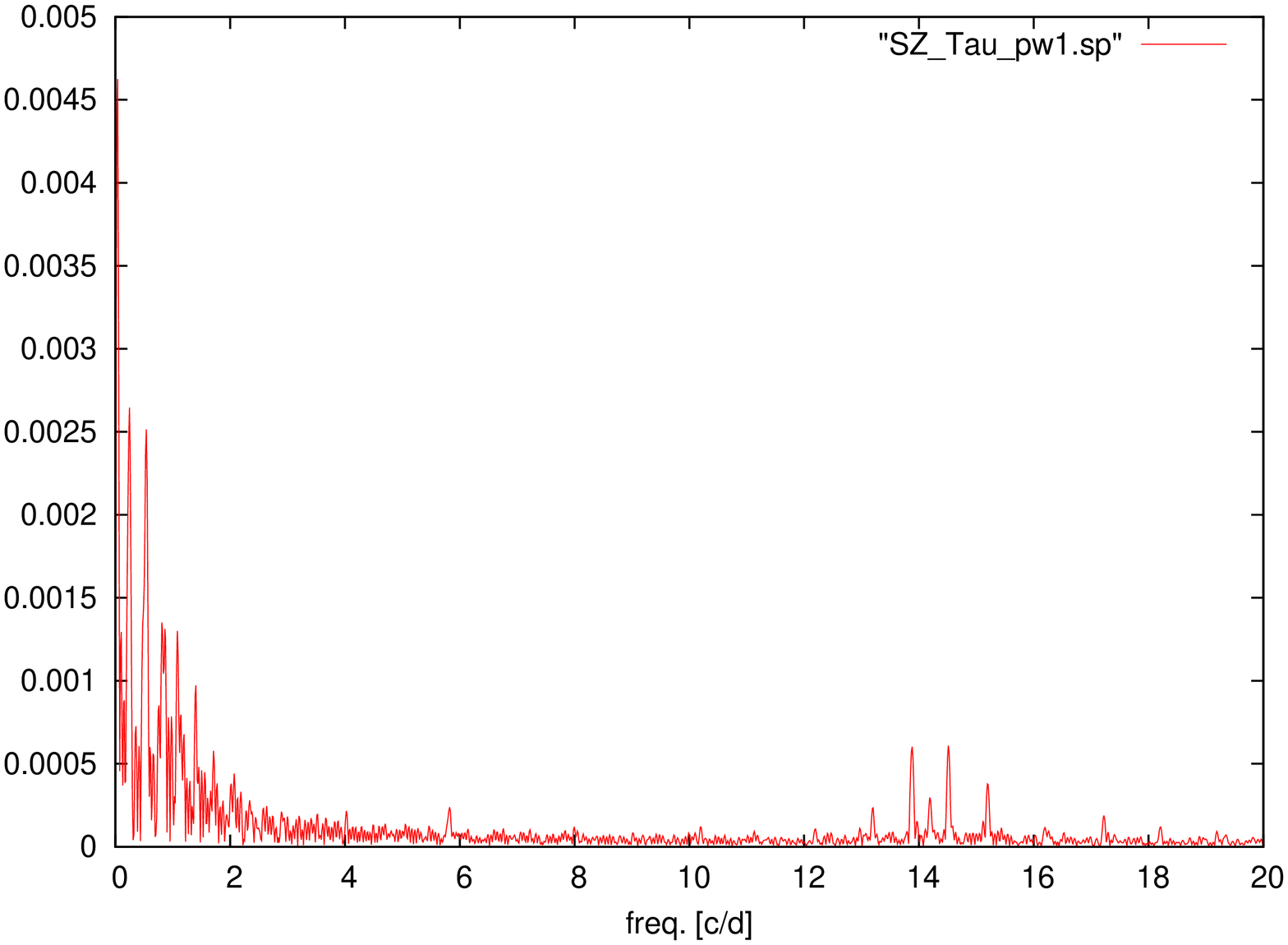}} 
%\end{subfigure}%
%\vspace*{.5 cm}
 \caption{The Fourier spectra of the {\it MOST} data (amplitude in
   mag. as a function of frequency in cycles/day)  of RT Aur (left) and SZ
 Tau (right). The data for both stars have been prewhitened by a
 Fourier series with 10 terms.  In both stars the instrumental signature is seen at
 about 14~d$^{-1}$. For RT Aur, little power is left at low frequencies.
 For SZ Tau, power still remains in a complicated set of low
 frequencies.   }
   \label{fig4}
\end{center}
\end{figure}

The data from the two stars was subjected to a number of comparisons.
In Fig.~\ref{fig4}, the Fourier spectra are shown after pre-whitening for a
Fourier fit of 10 terms.  The instrumental signal is seen at the
orbital frequency ($\simeq$14 d$^{-1}$).  The Fourier series describes
the fundamental mode pulsator (RT Aur) well and little power is left
at low frequencies.  In SZ Tau, on the other hand, power still remains
in a complicated set of low frequencies, indicating that the single
periodicity does not fully describe the pulsation. 

\begin{figure}
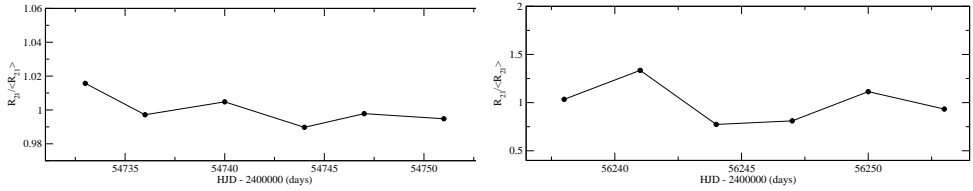

%\begin{subfigure}[b]{0.5\textwidth}
% \vspace*{-2.0 cm}
\begin{center}
\subfigure{\includegraphics[width=2.5in]{evans.fig5a.eps}}
% \end{subfigure}%
% \vspace*{-1.0 cm}
%\begin{subfigure}[b]{0.5\textwidth}
\subfigure{\includegraphics[width=2.5in]{evans.fig5b.eps}}
%\end{subfigure}%
\vspace*{2.5cm}
 \caption{Representative Fourier parameters for the {\it MOST} data
( $R_{21}/\langle R_{21}\rangle$ as a function of HJD - 2,400,000)
   for each of the six cycles observed, with RT Aur on the left and SZ
   Tau on the right.  We estimate that the
   $R_{21}/\langle R_{21}\rangle$ 
   parameter varies by 3\% in RT Aur and 45\% in SZ Tau. }
   \label{fig5}
\end{center}
\end{figure}

The Fourier parameters themselves were compared for the two stars.  As
an example, Fig.~\ref{fig5} shows the amplitude ratio $R_{21}$ changes for the
six cycles observed for each of the stars.  For RT Aur, the variation
is estimated to be 3\%.  For SZ Tau the variation is much larger
(45\%). The variation  in Fig. 5 (right) is typical of the variation
in other  Fourier coefficients of the
overtone Cepheid.  

Thus the {\it MOST} observations show that the overtone pulsator SZ Tau has
a larger variability in the parameters we examined, or a larger
instability in the pulsation than the fundamental mode Cepheid.  

\section{ Summary}
We present here a brief summary of the period changes or possible
period changes in classical Cepheids and the characteristics of
pulsation which they suggest.  Most probably the $\dot{P}$ which we
observe results from a combination of causes.  

\vspace*{1cm}

\vspace*{.5cm}

$\bullet$  Evolution through the instability strip:  This would have 
$\dot{P}$ in one direction (at least for a century of observations).

$\bullet$ Light-time effects in binaries: $\dot{P}$ would be cyclic
but with a long period.  (Known Cepheid binaries have orbital periods
longer than 1 year in the Milky Way.)

$\bullet$ Mass loss:  $\dot{P}$ would be in one direction,

$\bullet$ Star spots and rotation:  $\dot{P}$ would be cyclic and
roughly periodic as spots come and go.

The two phenomena  discussed here have the following characteristics:
\begin{itemize}
\item[--] Flickering $\dot{P}$ ({\it Kepler}):  It is cyclic and reasonably short
term.  It could be due either to a pulsation phenomenon or star spots.
\item[--] Instability in overtones ({\it MOST}):  This seems to be most probably
pulsation related.   
\end{itemize}

\begin{acknowledgments}
Financial Support was provided by CXC NASA Contract NAS8-03060 (NRE), 
ESTEC Contract No. 4000106398/12/NL/KML (LS), European Community's 
Seventh Framework Programme  (FP7/2007-2013) under Grant Agreement No. 269194
(IRSES/ASK) (RS, AD), the Hungarian OTKA grant K83790, 
the KTIA URKUT 10-1-2011-0019 grant,  
J\'anos Bolyai Research Scholarship of the Hungarian Academy of
Sciences (RS, AD), the Lend\"ulet-2009 Young Researchers Program of the
Hungarian Academy of Sciences (AD), and IAU travel grant (RS).
\end{acknowledgments}

\end{document}